\documentclass[11pt]{article}

\usepackage{amsmath,amssymb,color,verbatim}
\usepackage{pstricks}
\usepackage{graphicx}

\newcommand{\bnabla}{\boldsymbol{\nabla}}

\newcommand{\rem}[1]{}
\DeclareMathAlphabet{\mathbi}{OML}{cmm}{b}{it} 
\newcommand{\bx}{\mathbi{x}}

\newcommand{\bel}{\begin{equation}\label}
\newcommand{\ee}{\end{equation}}
\newcommand{\ben}{\begin{enumerate}}
\newcommand{\een}{\end{enumerate}}
\newcommand{\bde}{\begin{description}}
\newcommand{\ede}{\end{description}}
\newcommand{\bit}{\begin{itemize}}
\newcommand{\eit}{\end{itemize}}
\newcommand{\bc}{\begin{center}}
\newcommand{\ec}{\end{center}}

\newcommand{\bdB}{\mbox{\boldmath$\mathcal{B}$}}

\newcommand{\bu}{\mathbi{u}}

\newcommand{\bom}{\mbox{\boldmath$\omega$}}

\newcommand{\beq}{\begin{eqnarray}\label} 
\newcommand{\eeq}{\end{eqnarray}}

\newcommand{\non}{\nonumber}

\makeatletter
\rem{\@addtoreset{equation}{section}

\makeatother}

\begin{document}
\sf
\bc
\textbf{\Large A new diagnostic for the relative accuracy of Euler codes}
\par\vspace{7mm}
\textbf{\large C. R. Doering}
\par\vspace{1mm}
Department of Mathematics, Physics, and Center for the Study of Complex Systems,\\
University of Michigan, Ann Arbor, Michigan, MI 48109-1043
\par\vspace{1mm}
and
\par\vspace{1mm}
\textbf{\large J. D. Gibbon and D. D. Holm}
\par\vspace{1mm}
Department of Mathematics, Imperial College London SW7 2AZ, UK
\par\vspace{1mm}
{\small email\,: doering@umich.edu\,; j.d.gibbon@ic.ac.uk and d.holm@ic.ac.uk}
\ec
\begin{abstract}\noindent
{\small 
A procedure is suggested for testing the resolution and comparing the relative accuracy of numerical 
schemes for integration of the incompressible Euler equations.}
\end{abstract}

\vspace{2mm}
There is great interest in the behaviour of solutions of the three-dimensional incompressible 
Euler equations \cite{EE250}. An important open question is whether the vorticity field can 
develop a singularity in finite time; see, e.g., the review and literature cited in \cite{JDG250}.
A central challenge to the computational investigation of this question is the inevitable loss of 
accuracy of any numerical scheme for singular, or near singular, solutions. We do not attempt to 
answer this question\,; rather we propose a procedure for testing the accuracy and comparing the 
precision of numerical schemes used to integrate the incompressible Euler equations. 

The Euler equations for the velocity field $\bu$ ($\mbox{div}\,\bu = 0$) are
\bel{ee1}
\frac{D\bu}{Dt} = -\bnabla p\,,\qquad\qquad\frac{D~}{Dt}= \partial_{\,t} + \bu\cdot\bnabla\,,
\ee
and for the vorticity $\bom = \mbox{curl}\,\bu$,
\bel{ee2}
\partial_{\,t}\,\bom = \mbox{curl}\,{(\bu\times\bom)}\qquad\mbox{or}\qquad
\frac{D\bom}{Dt} = \bom\cdot\bnabla\bu\,.
\ee
The new feature of the proposed test is to introduce a passive tracer concentration $\theta(\bx,\,t)$
whose initial data are under the investigator's control, satisfying 
\bel{theta1}
\frac{D\theta}{Dt} = 0\,.
\ee
The introduction of $\theta$ allows us to define a variable $q=\bom\cdot\bnabla\theta$ \cite{gfd}.
An application of Ertel's theorem produces the well-known result
\bel{q1}
\frac{Dq}{Dt} = 0\,.
\ee
Now define
\bel{Bdef1}
\bdB = \bnabla q \times\bnabla \theta
\ee
which satisfies $\mbox{div}\,\bdB = 0$ and comes endowed with initial conditions inherited from 
those for $\bu$ and $\theta$. The fact is (for the proof see the Appendix) that $\bdB$ satisfies 
\cite{GH10a}
\bel{B1}
\partial_{\,t}\,\bdB = \mbox{curl}\,{(\bu\times\bdB)}\qquad\mbox{or}\qquad
\frac{D\bdB}{Dt} = \bdB\cdot\bnabla\bu\,.
\ee
\par\smallskip
The vector field $\bdB$ contains information on $\bom$, $\bnabla\bom$, $\bnabla\theta$ and $\bnabla^{2}\theta$
embedded in the combination in (\ref{Bdef1}). It evolves in the same way as $\bom$ and so is subjected to 
similar (potentially tortuous) stretching and folding processes. It can, however, be evaluated at any 
particular time $t$ in several distinct ways\,: it can be evaluated from the result of the evolution in 
(\ref{B1}) at time $t$, or it can be computed from its definition using $\bu$ and $\theta$ evolved up to 
and evaluated at time $t$. The degree to which these distinct evaluations agree or disagree provides a 
quantitative gauge of the accuracy of the numerical computation. It is not clear that there is a natural 
scale for the inevitable discrepancies produced in any particular computation. However, this procedure 
produces a precise diagnostic quantity that, given identical initial data, can be directly compared 
side-by-side for different numerical computations to evaluate their {\em relative} accuracy. Here is 
the suggested test\,:
\ben
\item Choose initial data for $\bu$ and $\theta$, thereby fixing initial data for $q$ and $\bdB$. 

\item Evolve $\bu$ and simultaneously solve $D\theta/Dt = 0$, $Dq/Dt = 0$ and $D\bdB/Dt = \bdB\cdot\bnabla\bu$.

\item Test the resolution at any time $t>0$ by constructing $q_{1}(\cdot\,,\,t) =
\bom(\cdot\,,\,t)\cdot\bnabla\theta(\cdot\,,\,t)$ and then\,:

\ben
\item compare the solution for $\bdB(\cdot,\,t)$ obtained from solving $D\bdB/Dt = \bdB\cdot\bnabla\bu$ with
$\bdB_{1}(\cdot\,,\,t) = \bnabla q_{1}(\cdot\,,\,t) \times \bnabla \theta(\cdot\,,\,t)$

\item and, furthermore, compare these with
$\bdB_{2}(\cdot\,,\,t) = \bnabla q(\cdot\,,\,t) \times \bnabla\theta(\cdot\,,\,t)$ 
where $q(\cdot\,,\,t)$ is the evolved solution of $Dq/Dt=0$.

\een
\item For fixed initial data for $\bu$ this procedure may be implemented for a variety of ``markers''
$\theta_{n}(\cdot\,,\,t)$ evolving from distinct initial data $\theta_{n}(\cdot\,,\,0)$ to diagnose
the numerical accuracy in different regions of the flow. 
\een
Because $\bdB$ contains $\bnabla\bom$, comparing the different computations of $\bdB,~\bdB_{1}$
and $\bdB_{2}$ tests the accuracy of the computation of some of the small scale
structures in the flow.
\par\bigskip\noindent
\textbf{\large Appendix}
\par\smallskip\noindent
Using conventional vector identities, the proof of (\ref{B1}) is (see \cite{GH10a})\,: 
\beq{stretch}
\bdB_{t} &=& (\nabla q)_{t}\times(\nabla\theta) + (\nabla q)\times(\nabla\theta)_{t}\non\\
&=& -\nabla\big(\bu\cdot\nabla q\big)\times(\nabla\theta)
- (\nabla q)\times\left[\nabla(\bu\cdot\nabla\theta)\right]\non\\
&=& -\left\{\bu\cdot\nabla(\nabla q) + (\nabla q)\cdot\nabla\bu 
+ (\nabla q)\times\bom\right\}\times(\nabla\theta)\non\\
&\quad& - (\nabla q)\times\left\{\bu\cdot\nabla(\nabla\theta) + 
(\nabla\theta)\cdot\nabla\bu + (\nabla\theta)\times\bom\right\}\non\\
&=& - \bu\cdot\nabla\bdB + (\nabla q)(\bom\cdot\nabla\theta) - (\nabla\theta)(\bom\cdot\nabla q)\non\\
&\quad& + (\nabla\theta)\times(\nabla q\cdot\nabla\bu) - (\nabla q)\times(\nabla\theta\cdot\nabla\bu)\non\\
&=& \mbox{curl}\,(\bu\times\bdB)
\eeq


\begin{thebibliography}{plain}\itemsep -1mm

\bibitem{EE250} Proc. Conf. \textit{Euler Equations\,: 250 Years On} (Aussois, France, June 18th-23rd, 2007), 
edited by Gregory Eyink, Uriel Frisch, Ren\'e Moreau and Andrei Sobolevskii, Physica D \textbf{237},
1825--2250 (2008).

\bibitem{JDG250}  J. D. Gibbon, \textit{The three-dimensional Euler equations\,: Where do we stand?}
Physica D \textbf{237}, 1894–-1904 (2008).

\bibitem{gfd} In geophysical fluid dynamics, when $\theta$ is the potential temperature, $q$ is known as
the potential vorticity.

\bibitem{GH10a} For discussion of the properties and origin of the vector $\bdB$, see J. D. Gibbon and
D. D. Holm, \textit{The dynamics of the gradient of potential vorticity}, arXiv:0911.1476v3 [nlin.CD]

\end{thebibliography}
\end{document}